\newcommand{\diracslash}[1]{#1\llap{/\kern2pt}}

\newcommand{\be}{\begin{equation}}
\newcommand{\ee}{\end{equation}}
\newcommand{\bea}{\begin{eqnarray}}
\newcommand{\eea}{\end{eqnarray}}
\newcommand{\ba}[1]{\begin{array}{#1}}
\newcommand{\ea}{\end{array}}

\newcommand{\bt}{\begin{tabular}}
\newcommand{\et}{\end{tabular}}

\newcommand{\beas}{\begin{eqnarray*}}
\newcommand{\eeas}{\end{eqnarray*}}

\documentclass[preprint,prd,aps,floats,nofootinbib,floatfix]{revtex4}

\DeclareSymbolFont{rsfs}{U}{rsfs}{m}{n}
\DeclareSymbolFontAlphabet{\mathrsfs}{rsfs}
\usepackage{amsmath}
\usepackage{slashed}
\usepackage{graphicx}
\usepackage{cleveref}
\usepackage{epstopdf}
\usepackage{xcolor}

\begin{document}

\title{Magnetic Moments of Octet Baryons in Hot and Dense Nuclear Matter}

 \author{Harpreet Singh}
\email{harpreetmscdav@gmail.com}
\affiliation{Department of Physics, Dr. B R Ambedkar National Institute of Technology Jalandhar, 
 Jalandhar -- 144011, Punjab, India}
 \author{Arvind Kumar}
\email{iitd.arvind@gmail.com, kumara@nitj.ac.in}
\affiliation{Department of Physics, Dr. B R Ambedkar National Institute of Technology Jalandhar, 
 Jalandhar -- 144011, Punjab, India}
  \author{Harleen Dahiya}
\email{dahiyah@nitj.ac.in}
\affiliation{Department of Physics, Dr. B R Ambedkar National Institute of Technology Jalandhar, 
 Jalandhar -- 144011, Punjab, India}

\def\be{\begin{equation}}
\def\ee{\end{equation}}
\def\bearr{\begin{eqnarray}}
\def\eearr{\end{eqnarray}}
\def\zbf#1{{\bf {#1}}}
\def\bfm#1{\mbox{\boldmath $#1$}}
\def\hf{\frac{1}{2}}
\def\kp{\zbf k+\frac{\zbf q}{2}}
\def\km{-\zbf k+\frac{\zbf q}{2}}
\def\hwo{\hat\omega_1}
\def\hwt{\hat\omega_2}

\begin{abstract}

We have calculated the in-medium magnetic moments of octet baryons in the presence of hot and dense symmetric nuclear matter. Effective magnetic moments of baryons have been derived from medium modified quark masses within chiral SU(3) quark mean field model. 
 Further, for better insight of medium modification of baryonic magnetic moments, we have considered the explicit contributions from the valence quarks, sea quarks as well as sea orbital angular momentum of sea quarks. These effects have been successful in giving the description of baryonic magnetic moments in vacuum.
  The magnetic moments of baryons are found to vary significantly as a function of density of nuclear medium.
\end{abstract}

\maketitle

\section{Introduction}
The study of in-medium properties of octet and decuplet baryons is of great importance in the present era. Heavy-ion collision experiments at various experimental facilities such as LHC at CERN \cite{beringer}, FSI at EMC \cite{ashman}, CBM at FAIR \cite{friese}, etc., are focused at the study of matter in the free space as well as in the presence of medium. The major goal of modern hadron accelerator facilities is to investigate the structure of hadrons by scattering experiments at large momentum transfer typically $1\text{GeV}/c^{2}$  and beyond, so as to map out various internal charge distributions underlying quark and gluon degrees of freedom. The main objectives of the heavy-ion collision facilities are to study the properties of hadrons in hadronic matter, chiral symmetry restoration at high temperature and density of medium, de-confinement phase from hadrons to QGP and to determine the equation of state for the hadronic matter at high density \cite{heuser,atlas,peter}. The experiments at various facilities, besides the data, require theoretical insight into the hadronic properties (such as magnetic moment, charge radii, electromagnetic form factors etc.) as well.

The magnetic moment of the particle plays an important role in the study of structure of matter at the sub-nuclear level as it largely depends upon its structure and structure parameters. Theoretically, the magnetic moments of octet as well as decuplet baryons have been extensively studied in the free space \cite{felix,wrb,hack,jg,jun,lks,ss}. Constituent quark model studies proposed that the baryonic magnetic moments can be calculated by summing the magnetic moments of constituent quarks \cite{cheng,cqm}. However, the values so obtained differ from those obtained experimentally.
         
Magnetic moments of the octet baryons have been calculated from the structure parameters of baryons such as electromagnetic form factors \cite{electro}. They are derived from the magnetic form factor $G_m(Q^2)$ at $Q^2=0$ (where $Q^2$ is squared four momentum of baryon) \cite{emprop}. They have also been extrapolated from the study of charge radii \cite{ryu2} and medium modified masses of the baryons \cite{string} as well. Covariant baryon chiral perturbation theory \cite{chp1,chp2} 
 has been extensively used to study octet baryon magnetic moments using the idea of SU(3) symmetry breaking and it has been shown that in the low energy regime the chiral expansion of octet baryon magnetic moments is possible if one consider the correction terms such as loop corrections and decuplet  degree of freedom to be small 
\cite{ref2,ref3,ref4,ref6,ref7,ref8,ref9}. However, in order to gain deeper insight of underlying quark dynamics, it is useful to consider the individual quark contribution to baryonic magnetic moments. MIT bag model provided a useful way to calculate baryonic magnetic moments 
considering the constituent quarks to be non-interacting \cite{electro}. Later on, weak coupling between the constituent quarks was proposed to include interactions of quarks in the baryons \cite{zandy}. The observed ratio of contribution from $u$-quark to the contribution from $d$-quark in the calculation of the total magnetic moment of nucleon as calculated by this approach can only be justified by considering dynamical quark masses \cite{emprop}.  Thus, one has to consider constituent quark masses in place of current quark masses to study the baryonic magnetic moments including quark dynamics. This fact gives a strong evidence of the presence of relativistic and gluon effects which are not accounted for in conventional quark models. 
        
Beside the free space calculations of baryonic structural properties (such as magnetic moments), the medium modification of these properties has always been an interesting aspect of QCD studies. Deep inelastic muon-nucleus scattering experiment at EMC has indicated that nucleon properties in nuclear medium can be different from their vacuum values \cite{aubert}. Similarly, magnetic moment of proton in $^{12}\text{C}$ seemed to be enhanced by about $25\%$ in nuclear medium as compared to its value in free space \cite{mulders}. 

Theoretical models for nuclear matter, such as Walecka model \cite{wal}, sigma model \cite{sig}, non-linear sigma model \cite{wang2}, Zimanyi and Moszkowski model \cite{song}, cloudy bag model \cite{mil}, Nambu-Jona-Lasinio (NJL) model \cite{bub} etc., have successfully  explained several properties of nuclear matter \cite{Gustavo,rau}. The key to success of models like NJL model in explaining the low energy baryonic dynamics is the assumption of hadrons having chiral quarks and interaction between the constituent quarks \cite{hatsuda}.
For better understanding of baryon properties in quark degrees of freedom, chiral quark models such as quark meson coupling (QMC) model were developed in similar lines as NJL model and cloudy bag model \cite{wang}. 
Medium modification of magnetic moments of octet baryons have been calculated using QMC model \cite{ryu3}. QMC model has been used at finite temperature and baryonic density and medium modification of magnetic moments of baryons have been derived through medium modification of bag radius. The results were quite close to the experimental data for the vacuum values. 

 In the present work we have used chiral SU(3) quark mean field model to calculate the  in-medium magnetic moments of baryon octet at finite temperature and density of the medium through the medium modification of baryon masses. The relation of baryon magnetic moments and corresponding effective quark masses have been derived in the analysis of hyperon static properties \cite{lipkin}. We will follow similar relations to obtain medium modified values of magnetic moments of baryons. Chiral SU(3) quark mean field model (CQMF) \cite{wang,wang2,wang3,wang4} has been extended from quark mean field model \cite{toki}, which is based on QMC model approach. In this model, mean field approximation is used, which uses classical expectation values in place of quantum field operators \cite{papag}. The quarks are assumed to be constituent quarks, which are confined in the baryons by confining potential. Finite nuclei properties have been studied in this model and reasonably good results have been obtained \cite{wang}. Within CQMF model, the in-medium masses of quarks and hence, baryons are calculated through the medium modification of scalar iso-scalar fields $\sigma$ and $\zeta$ and the scalar dilaton field $\chi$ \cite{wang,wang2,wang3,wang4}. 

 Beside the interaction of scalar meson fields, some entities having character of Goldstone boson (GB) play a major role in the interaction of quarks and their magnetic moments \cite{thaler}. If we assume Goldstone bosons in the interior of hadrons, we will have different propagation properties of the states \cite{cheng}. Spin dependent features of hadronic spectrum can be successfully explained by considering internal GB exchange between the quarks. Further, the significant spin-orbit coupling contribution can also be accounted for by this approach. Beside this, the violation of Gottfried sum rule leads to the isospin asymmetric sea quark in baryons, and sea quark contributions should also be considered in magnetic moments of baryons \cite{rfw,johan,simon,hack}. In this work we have considered the GB exchange in the interior of baryon, and also, we have considered the contribution from sea quark. These two effects can further modify the effective magnetic moments. 
           
           The outline of the paper is as follows: In section \ref{masscalcua} we will apply CQMF model to find the effective quark masses at finite temperature and density of nuclear medium, and hence, calculate effective baryon octet masses. We will discuss the effect of valence quarks, sea quarks and orbital angular momentum of sea quarks on the magnetic moments of baryons in section \ref{masscalcub}. The section \ref{results} is devoted to numerical calculations and results. Section \ref{summ} includes the summary of present work.
     
\section{Model} 
\subsection{Chiral $\textbf{SU(3)}$ Quark Mean Field Model for Quark Masses} \label{masscalcua}
To study the  structure of hadrons in chiral limit and explore it in quark degrees of freedom, the quarks are divided into two parts, left-handed `$q_L$' and right-handed `$q_R$'. Under $\text{SU(3)}_L \times \text{SU(3)}_R$ transformation, the corresponding transformations for the left and right handed quarks are
\begin{equation}
q_{L} \rightarrow q_{L}^{\prime }\,=\,L\,q_{L},~~~~~
q_{R} \rightarrow q_{R}^{\prime }\,=\,R\,q_{R},\,
\end{equation} 
where `$L$' and `$R$' are global $\text{SU(3)}_L \times \text{SU(3)}_R$ transformations given as
\begin{equation}
L(\alpha_L)=\text{exp}\left[ i\sum_{a=0}^{8} {\alpha}_L^a {\lambda}_{La}\right] ,\,~R(\alpha_R)=\text{exp}\left[ i\sum_{b=0}^{8} {\alpha}_R^b {\lambda}_{Rb}\right] ,
\end{equation}
$\alpha_L$ and $\alpha_R$ represent space-time independent parameters with indicies $(a=0,..,8)$ and $(b=0,..,8)$. $\lambda_L$ and $\lambda_R$ are Gell-Mann matrices  written as
\begin{equation}
\lambda_L=\lambda\frac{( 1-\gamma_5)}{2}, \,~~~~~~~~
\lambda_R=\lambda\frac{( 1+\gamma_5)}{2}.
\end{equation}
The nonents of spin-0 scalar ($\Sigma$) and pseudoscalar ($\Pi$) mesons can be written in compact form using Gell-Mann matrices as
\begin{equation}
M(M^{\dagger})=\Sigma \pm i\Pi =\frac{1}{\sqrt{2}}\sum_{a=0}^{8}
\left( s_{a}\pm i p _{a}\right) \lambda _{a},
\end{equation}
where $\lambda _{a}$ are Gell-Mann matrices with $\lambda _{0}=\sqrt{\frac{2}{3}}I$, $s_{a}$ and $p_{a}$ are the nonets of scalar and pseudoscalar
mesons, respectively. The plus and minus signs are for  $M$ and $M^{\dagger}$, respectively, which transform under chiral $\text{SU(3)}$ transformation as 
\begin{equation}
M\rightarrow M^{\prime }=LMR^{\dagger},
\end{equation}
\begin{equation}
 M^{\dagger}\rightarrow
M^{{\dagger}^{\prime }}=RM^{\dagger}L^{\dagger}.
\end{equation}
In the similar way, spin-1 mesons are defined by 
\begin{equation}
l_{\mu }(r_{\mu })=\frac{1}{2}\left( V_{\mu }\pm A_{\mu }\right)
= \frac{1}{2\sqrt{2}}\sum_{a=0}^{8}\left( v_{\mu}^{a}\pm a^{a}_{\mu}
\right) \lambda_{a}.
\end{equation}
where $v_{\mu}^a$ and $a_{\mu}^a$ are nonets of vector and pseudovector mesons.
The alternative plus and minus signs are for $l_{\mu}$ and $r_{\mu}$ respectively, and will transform under chiral $\text{SU(3)}$ transformation as 
\begin{equation}
l_{\mu}\rightarrow l_{\mu }^{\prime }=Ll_{\mu }L^{\dagger},
\end{equation}
\begin{equation}
r_{\mu}\rightarrow r_{\mu }^{\prime }=Rr_{\mu }R^{\dagger}.
\end{equation}
 The physical states for scalar and vector mesons are explicitly represented as
\begin{equation}
\Sigma = \frac1{\sqrt{2}}\sum_{a=0}^8 s_a \, \lambda_a=\left(
\begin{array}{lcr}
\frac1{\sqrt{2}}\left(\sigma+a_0^0\right) & a_0^{+} & \kappa^{*+} \\
a_0^- & \frac1{\sqrt{2}}\left(\sigma-a_0^0\right) & \kappa^{*0} \\
\kappa^{*-} & \bar{\kappa}^{*0} & \zeta
\end{array}
\right),
\end{equation}
and
\begin{equation}
V_\mu = \frac1{\sqrt{2}}\sum_{a=0}^8 v_\mu^a \, \lambda_a=\left(
\begin{array}{lcr}
\frac1{\sqrt{2}}\left(\omega_\mu+\rho_\mu^0\right)
& \rho_\mu^+ & K_\mu^{*+}\\
\rho_\mu^- & \frac1{\sqrt{2}}\left(\omega_\mu-\rho_\mu^0\right)
& K_\mu^{*0}\\
K_\mu^{*-} & \bar{K}_\mu^{*0} & \phi_\mu
\end{array}
\right),
\,
\end{equation}
respectively. 
In a similar manner, we can write pseudoscalar nonet ($\Pi$) and pseudovector nonet ($A_{\mu}$). The total effective Lagrangian density in chiral $\text{SU(3)}$ quark mean field model is written as
\begin{equation}
{\cal L}_{{\rm eff}} \, = \, {\cal L}_{q0} \, + \, {\cal L}_{qm}
\, + \,
{\cal L}_{\Sigma\Sigma} \,+\, {\cal L}_{VV} \,+\, {\cal L}_{\chi SB}\,
+ \, {\cal L}_{\Delta m} \, + \, {\cal L}_{c}, \label{totallag}
\end{equation}
where ${\cal L}_{q0} =\bar q \, i\gamma^\mu \partial_\mu \, q$ represents the free part of massless quarks, ${\cal L}_{qm}$ is the chiral SU(3)-invariant quark-meson interaction term and is written as
\begin{eqnarray}
{\cal L}_{qm} & = & g_s\left(\bar{\Psi}_LM\Psi_R+\bar{\Psi}_RM^+\Psi_L\right) \nonumber \\
& - & g_v\left(\bar{\Psi}_L\gamma^\mu l_\mu\Psi_L+\bar{\Psi}_R\gamma^\mu
r_\mu\Psi_R\right)  \nonumber \\
& = & \frac{g_s}{\sqrt{2}}\bar{\Psi}\left(\sum_{a=0}^8 s_a\lambda_a
 +  i \gamma^5 \sum_{a=0}^8 p_a\lambda_a
\right)\Psi \nonumber \\
&-& \frac{g_v}{2\sqrt{2}}
\bar{\Psi}\left( \gamma^\mu \sum_{a=0}^8
 v_\mu^a\lambda_a 
 -  \gamma^\mu\gamma^5 \sum_{a=0}^8
a_\mu^a\lambda_a\right)\Psi,
\nonumber \\
\end{eqnarray}
where $\Psi=\left(\begin{array}{lcr}
u
& \\
d
& \\
s
\end{array}\right)
$.  
The chiral-invariant scalar and vector meson self interaction terms ${\cal L}_{\Sigma\Sigma}$ and ${\cal L}_{VV}$, within mean field approximation \cite{wang3} are written as
\begin{eqnarray}
{\cal L}_{\Sigma\Sigma} &=& -\frac{1}{2} \, k_0\chi^2
\left(\sigma^2+\zeta^2\right)+k_1 \left(\sigma^2+\zeta^2\right)^2
+k_2\left(\frac{\sigma^4}2 
+ \zeta^4\right) \nonumber \\
&+&k_3\chi\sigma^2\zeta
 -k_4\chi^4-\frac14\chi^4 {\rm ln}\frac{\chi^4}{\chi_0^4} 
 +\frac{\xi} 3\chi^4 {\rm ln}\frac{\sigma^2\zeta}{\sigma_0^2\zeta_0}, \label{scalar0}
\end{eqnarray}    
and
\begin{equation}
{\cal L}_{VV}=\frac{1}{2} \, \frac{\chi^2}{\chi_0^2} \left(
m_\omega^2\omega^2\right)+g_4
\omega^4, \label{vector}
\end{equation}
respectively.
 The constants $k_0, k_1, k_2, k_3$ and $k_4$ appearing in equation (\ref{scalar0}) are determined using $\pi$ meson mass ($m_{\pi}$), $K$ meson mass ($m_K$) and the average mass of $\eta$ and $\eta^{'}$ mesons \cite{wang}. The other parameters, i.e., $\xi$, vacuum value of dilaton field, $\chi_0$, and, the coupling constant $g_4$, are chosen so as to fit effective nucleon mass reasonably. Further, the value of parameter `$\xi$' originating from logarithmic term used in scalar meson self interaction Lagrangian density can be obtained using QCD $\beta$-function at one loop level, for three colors and three flavors \cite{papag}. The Lagrangian density ${\cal L}_{\chi SB}$ in equation (\ref{totallag}) is introduced to incorporate non-vanishing pesudoscalar meson masses and it satisfies the partial conserved axial-vector current relations for $\pi$ and $K$ mesons \cite{wang,wang3,wang4}. We have
\begin{equation}\label{L_SB}
{\cal L}_{\chi SB}=\frac{\chi^2}{\chi_0^2}\left[m_\pi^2F_\pi\sigma +
\left(
\sqrt{2} \, m_K^2F_K-\frac{m_\pi^2}{\sqrt{2}} F_\pi\right)\zeta\right],
\end{equation}
where $F_{\pi}$ and $F_K$ are pion and kaon decay constants, respectively. Masses of `$u$', `$d$' and `$s$' quarks are generated by the vacuum expectation values of $\sigma$ and $\zeta$ mesons scalar fields. In order to find constituent strange quark mass correctly, an additional mass term which would explicitly break the chiral symmetry is written in equation (\ref{totallag}). This term can be expressed as
\begin{equation}
{\cal L}_{\Delta m} = - {m}_{1} \bar \Psi S_1 \Psi,
\end{equation}
where $m_1$ is the additional mass term. The strange quark matrix operator $S_1$ is defined as
$S_1 \, = \, \frac{1}{3} \, \left(I - \lambda_8\sqrt{3}\right) =
{\rm diag}(0,0,1)$.
Thus, the relations for vacuum masses of quarks are 
\begin{equation}
\label{qvacmass}
m_u=m_d=-g_{\sigma}^q \sigma_0=-\frac{g_s}{\sqrt{2}}\sigma_0,
\hspace*{.3cm} \mbox{and} \hspace*{.3cm}
m_s=-g_{\zeta}^s \zeta_0 + m_1. 
\end{equation}
The values of coupling constant $g_s$ and additional mass term $m_1$ in equation (\ref{qvacmass}) can be calculated by taking $m_u=m_d=313$ MeV and $m_s=490$ MeV as vacuum masses of quarks. The interaction between the quarks and vector mesons leads to \cite{wang3} 
\begin{eqnarray}
\frac{g_s}{\sqrt{2}}
= g_\sigma^u = g_\sigma^d = 
\frac{1}{\sqrt{2}}g_\zeta^s,  \nonumber \\
~~~g_\sigma^s=g_\zeta^u = g_\zeta^d = 0, \nonumber \\
g_\omega^u = g_\omega^d = g_\omega^q, \nonumber \\
~~~~~~~~~~g_\omega^s  = 0 .
\end{eqnarray} 
Quarks are confined in baryons by confining scalar-vector potential as, given by \cite{wang3}  
\begin{equation}
\chi_{c}(r)=\frac14 k_{c} \, r^2(1+\gamma^0) \,.   \label{potential}
\end{equation}
The coupling constant $k_c$ is taken to be $100 \, \text{MeV}. \text{fm}^{-2}$.
Corresponding Lagrangian density is written as 
\begin{equation}
{\cal L}_{c} = -  \bar \Psi \chi_c \Psi.
\end{equation}
In order to investigate the properties of nuclear matter at finite temperature and density, we will use mean field approximation \cite{wang3}. The Dirac equation under the influence of meson mean field, for the quark field $\Psi_{qj}$ is given by 
\begin{equation}
\left[-i\vec{\alpha}\cdot\vec{\nabla}+\chi_c(r)+\beta m_q^*\right]
\Psi_{qj}=e_q^*\Psi_{qj}, \label{Dirac}
\end{equation}
where the subscripts $q$ and $j$ denote the quark $q$ ($q=u, d, s$)
in a baryon of type $j$ ($j=N, \Lambda, \Sigma, \Xi$)\,
and $\vec{\alpha}$\,, $\beta$\, are usual Dirac matrices.
The effective quark mass $m_{q}^*$ is defined as
\begin{equation}
m_q^*=-g_\sigma^q\sigma - g_\zeta^q\zeta+m_{q0}, \label{qmass}
\end{equation}
where $m_{q0}=m_1$ is zero for non-strange `$u$' and `$d$' quarks, whereas for strange `$s$' quark $m_{q0}=m_1=29$ MeV. Effective energy of particular quark under the influence of meson field is given as,
$e_q^*=e_q-g_\omega^i\omega-g_\phi^i\phi \,$ \cite{wang,wang3}. For the  confining potential defined by equation (\ref{potential}), the analytical expression for effective energy of quark $e_q^*$ will be
\begin{equation}
e_q^*=m_q^*+\frac{3\sqrt{k_c}}{\sqrt{2(e_q^*+m_q^*)}}.  \label{qenergy}
\end{equation} 
The effective mass of baryons can be calculated from the effective quark masses $m_q^*$, using the relation  
\begin{equation}
M_j^*=\sqrt{E_j^{*2}- <p_{j \, \text{cm}}^{*2}>}\,, \label{baryonmass}
\end{equation} 
where the effective energy of $j^{th}$ baryon in the nuclear medium is given as
\begin{equation} 
E_j^*=\sum_qn_{qj}e_q^*+E_{j \, \text{spin}}.    \label{energy}
\end{equation}
 Further, $E_{j \, \text{spin}}$ is the correction to baryon energy due to spin-spin interaction of constituent quarks and takes the values 
\begin{eqnarray}
E_{N \, \text{spin}} = - 477 \,\,\, {\rm MeV}\,,
~~E_{\Lambda \, \text{spin}} = - 756.9 \,\,\, {\rm MeV}\,,\nonumber \\
E_{\Sigma \, \text{spin}} =  - 531 \,\, {\rm MeV}\,, 
~~E_{\Xi \, \text{spin}} =  - 705 \,\, {\rm MeV}\, \nonumber.
\end{eqnarray}
These values are determined to fit the respective vacuum values of baryon masses. In equation (\ref{baryonmass}), $<p_{j \, \text{cm}}^{*2}>$ is the spurious center of mass motion \cite{barik1,barik2}.
 To study the equations of motion for mesons at finite temperature and density, we consider the thermopotential as
\begin{eqnarray}
\Omega &=&-\sum_{B = N\,, \Lambda\,, \Sigma\,, \Xi }
\frac{g_j k_{B}T}
{(2\pi)^3} \nonumber \\
&& \int_0^\infty d^3k\biggl\{{\rm ln}
\left( 1+e^{- [ E^{\ast}(k) - \nu_B ]/k_{B}T}\right) \nonumber \\
&+& {\rm ln}\left( 1+e^{- [ E^{\ast}(k)+\nu_B ]/k_{B}T}
\right) \biggr\} -{\cal L}_{M},  
\end{eqnarray}
where
\begin{equation}
{\cal L}_{M} \, = 
{\cal L}_{\Sigma\Sigma} \,+\, {\cal L}_{VV} \,+\, {\cal L}_{\chi SB}\,,
\end{equation} 
and $g_j$ is degeneracy of $j^{\text{th}}$ baryon ($g_{N, \Xi}=2$,
$g_\Lambda=1$, $g_\Sigma=3$) and $E^{\ast }(k)=\sqrt{M_j^{\ast 2}+k^{2}}$. We can relate the quantity $\nu_B$ to the chemical potential $\mu_B$ as \cite{wang,wang3,wang4}
\begin{equation}
\nu_B = \mu_B - g_{\omega}^j\omega .
\end{equation}
The equations of motion for scalar fields $\sigma$, $\zeta$, the dilaton field, $\chi$, and the vector field $\omega$ are calculated from thermodynamical potential and are respectively expressed as 
\begin{eqnarray}
\label{eq_sigma}
&&k_{0}\chi^{2}\sigma
-4k_{1}\left( \sigma^{2} \, + \, \zeta^{2}\right) \sigma \, - \,
2k_{2} \sigma^{3}\, - \, 2k_{3}\chi \sigma \zeta \, - \,
\frac{2\xi }{3\sigma }\chi^{4}+
\, \nonumber \\  &&\, \frac{\chi^{2}}{\chi _{0}^{2}}m_{\pi }^{2}F_{\pi } 
-\left( \frac{\chi }{\chi _{0}}\right)^{2}m_{\omega }\omega ^{2}
\frac{\partial m_{\omega }}{\partial \sigma }\, + \,
 \frac{\partial M_{N}^{\ast }}
{\partial \sigma } <\bar{\psi _{N}}\psi_{N}>=0,
\nonumber \\
\end{eqnarray}
\begin{eqnarray}
\label{zeta}
&&k_{0}\chi^{2}\zeta - 4k_{1}\left(\sigma^{2} \, + \,
\zeta ^{2}\right)
 \zeta \, - \, 4k_{2}\zeta ^{3} \, - \,
k_{3}\chi \sigma ^{2} \, - \, \frac{\xi }{3\zeta }\chi^{4} \, + \nonumber \\ 
&&\frac{\chi^{2}}{\chi _{0}^{2}} \left( \sqrt{2}m_{K}^{2}F_{K} \, - \,
\frac{1}{\sqrt{2}}m_{\pi }^{2}F_{\pi } \right)=0,
\nonumber \\
\end{eqnarray}
\begin{eqnarray}
\label{scalar1}
&&k_0\chi^2
\left(\sigma^2+\zeta^2\right)
-k_3\chi\sigma^2\zeta
 +\left(4k_4+1-4{\rm ln}\frac{\chi^4}{\chi_0^4} +
\frac{4\xi}
3{\rm ln}\frac{\sigma^2\zeta}{\sigma_0^2\zeta_0}\right)\chi^3 \nonumber \\
&&+\frac{2\chi}{\chi _{0}^{2}}\left[  m_{\pi }^{2}F_{\pi }\sigma+\left( \sqrt{2}m_{K}^{2}F_{K} \, - \,
\frac{1}{\sqrt{2}}m_{\pi }^{2}F_{\pi } \right)\zeta\right]- \frac{\chi}{\chi_0^2} 
m_\omega^2\omega^2 =0,
\nonumber \\
\end{eqnarray}
and
\begin{equation}
 \frac{\chi^2}{\chi_0^2} \left(
m_\omega^2\omega^2\right)+4g_4
\omega^3=g_\omega^N<{\psi _{N}}^{\dagger}\psi_{N}>. \label{vector}
\end{equation}
In equation (\ref{eq_sigma}), $<\bar{\psi_N}\psi_N>$ is the scalar density of nucleons and is  given by
\begin{eqnarray}
<\bar{\psi_N}\psi_N> =&&\frac{g_N \,}{2\pi^2} \,
\int_{0}^{{\infty}} dk \frac{k^2 M_j^\ast}{\sqrt{M_j^{\ast 2}+k^2}}\nonumber \\ 
&&\left[n_n(k)+ \bar n_n(k)+n_p
(k)+ \bar n_p(k)\right]. 
\end{eqnarray}
The number density of nucleons in equation (\ref{vector}) is given as
\begin{equation}
<{\psi_N}^{\dagger}\psi_N> =\frac{g_N \,}{2\pi^2} \,
\int_{0}^{{\infty}} dk k^2 \left[n_n(k)
+ \bar n_n(k)+n_p
(k)+ \bar n_p(k)\right] ,
\end{equation}
where, $n_n(k)$ and $n_p(k)$ are the neutron and proton distributions, and, $\bar n_n(k)$ and $\bar n_p(k)$ are the anti-neutron and anti-proton distributions, respectively and are defined as
\begin{equation}
n_{\tau}(k)=\lbrace \text{exp}\left[ \left( E^{\ast }(k)-\nu_{B} \right) /k_{B}T \right]+1\rbrace^{-1},~~~~~
\end{equation}
\begin{equation}
\bar n_{\tau}(k)=\lbrace \text{exp}\left[ \left( E^{\ast }(k)+\nu_{B} \right) /k_{B}T \right]+1\rbrace^{-1},~~(\tau= n,p).
\\
\end{equation}
The vacuum expectation values of meson fields $\sigma_0$ and $\zeta_0$ are constrained because of spontaneous breaking of chiral symmetry and are represented in terms of pion and kaon leptonic decay constants as
\begin{equation}
\sigma_0= -F_{\pi} ~~,~~~~~~~~~  \zeta_0= \frac{1}{\sqrt{2}}\left( F_{\pi}-2F_{K}\right). 
\end{equation} 
For $F_{\pi}=92.8$ MeV and $F_{K}=115$ MeV, the vacuum values of $\sigma$ and $\zeta$ fields are $\sigma_0=-92.8$ MeV and $\zeta_0=-96.5$ MeV respectively.
\subsection{Magnetic Moment of Baryons}
\label{masscalcub} 
So far we have used chiral $\text{SU(3)}$ quark mean field model  using effective Lagrangian density for the various interactions for calculating effective mass of constituent quarks. In order to calculate explicit contribution of valence and sea quark effect for the magnetic moment of baryons, we follow the idea of chiral quark model initiated by Weinberg \cite{sw} and developed by Manohar and Georgi \cite{am}. The model incorporate the idea of confinement and chiral symmetry breaking. The massless quarks acquire the mass through spontaneous breaking of chiral symmetry. The basic process in this approach is the emission of GB, which further splits into a $q\bar{q}$ pair,  e.g.,
\be
  q_{\pm} \rightarrow {\rm GB}^{0}
  + q^{'}_{\mp} \rightarrow  (q \bar q^{'})
  +q_{\mp}^{'}\,,                              \label{basic}
\ee
where $q \bar q^{'}  +q^{'}$ constitute the `sea quark'
  \cite{{chengsu3},{cheng1},{song},{johan}}. Within the QCD confinement scale and chiral symmetry breaking, the constituent quarks, octet of GBs and the weakly interacting gluons are the appropriate degrees of freedom \cite{aarti}. The effective Lagrangian in this region is given as
\be 
{\cal L}_{{\rm interaction}} = \bar{\psi}\left(i{\slashed D} + {\slashed V}\right)\psi +
i g_{A} \bar{\psi} {\slashed A}\gamma^{5}\psi + \cdots \,,
\ee
where $g_A$ is axial vector coupling constant. In the low energy limit gluonic degrees can be neglected. Hence, the above effective interaction Lagrangian with GBs and quarks in leading order is written as
\be {\cal
L}_{{\rm interaction}} = -\frac{g_{A}}{f_{\pi}} \bar{\psi} \partial_{\mu}
\Phi \gamma^{\mu} \gamma^{5} \psi \,, 
\ee 
using the Dirac's equation $(i \gamma^{\mu} \partial_{\mu} - m_q)q =0$, the effective Lagrangian describing interaction between quarks and a nonet of
GBs consisting of octet and a singlet and supressing all the space-time structure to lowest order can be expressed as

\be
{\cal L} = g_8 \bar q \Phi q\,,
\ee
with, 
\bea 
&&q =\left( \ba{c} u \\ d \\ s \ea \right) ,\nonumber \\
&&\resizebox{.9\hsize}{!}{ 
$\Phi = \left( \ba{ccc} \frac{\pi^o}{\sqrt 2}
+\varpi\frac{\eta}{\sqrt 6}+\tau\frac{\eta^{'}}{\sqrt 3} & \pi^+
  & \varepsilon K^+   \\
\pi^- & -\frac{\pi^o}{\sqrt 2} +\varpi \frac{\eta}{\sqrt 6}
+\tau\frac{\eta^{'}}{\sqrt 3}  &  \varepsilon K^o  \\
 \varepsilon K^-  &  \varepsilon \bar{K}^o  &  -\varpi \frac{2\eta}{\sqrt 6}
 +\tau\frac{\eta^{'}}{\sqrt 3} \ea \right).$} 
 \nonumber \\
 \eea
In above, $\varepsilon$, $\varpi$ are symmetry breaking parameters. Further, the parameter $\tau=g_1/g_8$, where  $g_1$ and $g_8$ are the coupling constants for the 
singlet and octet GBs, respectively. However, in accordance with New Muon Collaboration {\cite{nmc}} calculations we have used the value of $\tau$ obtained according to relation
\begin{equation}
\tau=-0.7-\frac{\varpi}{2} \label{zetadash}
\end{equation}

$\text{SU(3)}$ symmetry breaking is introduced by considering
$m_s > m_{u,d}$, as well as by considering
the masses of GBs to be nondegenerate
 $(m_{K,\eta} > m_{\pi})$ {\cite{{chengsu3},{cheng1},{song},{johan}}}. The octet baryon wave functions include singlet and triplet states and the gluon exchange forces  generates the mixing between them. Following the Cheng and Li mechanism \cite{cheng}, the magnetic moment of baryons, including the contributions from valence quarks, sea quarks and the orbital angular momentum of sea quark 
can be written as
 \be
 \mu\left( B\right)_{\text{total}}= \mu\left( B\right)_{\text{val}}+\mu\left( B\right)_{\text{sea}}+\mu\left( B\right)_{\text{orbital}}, \label{magtotal}
 \ee
 where $\mu\left( B\right)_{\text{val}}$ and  $\mu\left( B\right)_{\text{sea}}$ represent the contribution from valence and sea quarks, respectively. 
 The valence and sea contributions in terms of quark spin
polarizations can be written as 
\be
\mu(B)_{{\rm val}}=\sum_{q=u,d,s} {\Delta q_{{\rm val}}\mu_q} ~~~ 
{\rm and } ~~~
\mu(B)_{{\rm sea}}=\sum_{q=u,d,s} {\Delta q_{{\rm sea}}\mu_q}\,,    \label{mag}
\ee
where ${\Delta q}_{\rm val}$ and ${\Delta q}_{\rm sea}$ are the spin polarizations due to valence quarks and sea quark, respectively. Quark spin polarization is defined as 
\begin{equation}
{\Delta q}=q^{+}-q^{-}+\bar{q^+}-\bar{q^-},
\end{equation}
 where $q^+(\bar{q^+})$ and $q^-(\bar{q^-})$ is number of quarks (antiquarks) with spin up and down, respectively. The sum of ${\Delta q}$'s give total spin carried by  quarks. The spin structure of baryon is given as 
\begin{eqnarray} 
\widehat{B}=<B|\textit{N}|B>,  \nonumber 
\end{eqnarray}
where $\textit{N}$ is the number operator corresponding to different quark flavors with spins up and down and expressed as 
\begin{eqnarray} 
\textit{N}=n_{u^+}u^++n_{u^-}u^-+n_{d^+}d^++n_{d^-}d^-+n_{s^+}s^++n_{s^-}s^- , \nonumber 
\end{eqnarray}
with coefficient of $q^\pm$ giving the number of $q^\pm$ quarks. The  calculation of number of up and down quarks in a specific baryon have been explicitly done in Ref. \cite{har}.
  The sea quark polarization $\Delta q_{\rm sea}$ can be  expressed in terms of symmetry breaking parameters $\varepsilon$ and $\varpi$. For example, in case of proton $\Delta u_{\rm sea}$, $\Delta d_{\rm sea}$ and $\Delta s_{\rm sea}$ are defined as
 \begin{eqnarray}
 \Delta u_{\rm sea}=-\frac{a}{3}\left[7+4\varepsilon^{2}+\frac{4}{3}\varpi^{2}+\frac{8}{3}\tau^{2}\right], \nonumber \\
 \Delta d_{\rm sea}=-\frac{a}{3}\left[2-\varepsilon^{2}-\frac{1}{3}\varpi^{2}-\frac{2}{3}\tau^{2}\right], \nonumber \\
 \Delta s_{\rm sea}=-a\varepsilon^{2}, \label{seapol}
\end{eqnarray} 
respectively. The pion fluctuation parameter `$a$' is taken to be $0.1$, in the symmetric limit \cite{pion}.   
Similarly, $\Delta q_{\rm sea}$ is defined for other baryons in Ref. \cite{har,cheng}.  

The values of effective magnetic moment of constituent quark ($\mu_q$) can be calculated following the naive quark model formula given as 
$\mu_{\rm q}=\frac{e_{\rm q}}{2m_{\rm q}}$, where $m_{\rm q}$ and $e_{\rm q}$ are mass and electric charge of quark, respectively. This formula lacks consistency for calculation of magnetic moments of relativistically confined quarks \cite{har2}. Further, the non-relativistic quark momenta are required to be very small ($p^2_q<<(350 {\rm MeV})^2$) for quark masses in the range of $313$ MeV and more. Hence , in order to include quark confinement effect on magnetic moment \cite{har2,gupta} along with relativistic correction to quark magnetic moments (introduced in quarks by using medium modified quark masses obtained in chiral SU(3) quark mean field model, which considers quarks as Dirac particles), the mass term in the formula for quark magnetic moment is replaced by the expectation value of effective quark mass $\bar{m^{\rm B}_{\rm q}}$, which can be further expressed in terms of effective baryon mass following the formula 
\begin{eqnarray}
2\bar{m^{\rm B}_{\rm q}}=M_B^*+m_q+\Delta M \nonumber,
\end{eqnarray} 
where $M_B^*$ is effective mass of baryon, $m_q(\approx 0)$ is the current quark mass and $\Delta M$ is the confinement correction term \cite{har2}. 

Following the above formalism the equations 
to calculate effective magnetic moments `${\mu_q}$' of constituent quarks are now given as
\begin{equation}
 \mu_d =-\left(1-\frac{\Delta M}{M_B^*}\right),~~  \mu_s=-\frac{m_u^*}{m_s^*}\left(1-\frac{\Delta M}{M_B^*}\right),~~  \mu_u=-2\mu_d .\label{magandmass}         
\end{equation}     
The equation (\ref{magandmass}) are known mass adjusted magnetic moments of constituent quarks \cite{aarti}. $M_B^*$ is obtained in equation (\ref{baryonmass}). To include quark confinement effect it is replaced by $M_B^*+\Delta M$. $\Delta M$ being the difference between experimental vacuum mass of baryon ($M_{\rm vac}$) and the effective mass of baryon $M_B^*$,  i. e., $\Delta M=M_{\rm vac}-M_{B}^{*}$.   

The contribution from orbital angular momentum of sea quarks for the octet baryon of the type B(xxy) is given as
\begin{equation}
\mu(B(\text{xxy}))_{\rm orbit}=\Delta \text{x} \left[ \mu\left(\text{x}^+\rightarrow \right)\right]+ \Delta \text{y} \left[ \mu\left(\text{y}^+\rightarrow \right)\right],
\label{baryonxxy}
\end{equation}
and for the baryon of the type B(xyz) is 
\begin{eqnarray}
\mu(B(\text{xyz}))_{\rm orbit} &&= \Delta \text{x} \left[ \mu\left(\text{x}^+\rightarrow \right)\right]+ \Delta \text{y} \left[ \mu\left(\text{y}^+\rightarrow \right)\right] \nonumber \\ 
&&+ \Delta \text{z} \left[ \mu\left(\text{z}^+\rightarrow \right)\right].
\label{baryonxyz}
\end{eqnarray}
In equations (\ref{baryonxxy}) and (\ref{baryonxyz}), the symbols x, y and z correspond to any of the constituent quarks of the baryon, i.e., $u, d$ or $s$, and, $\Delta \text{x}$, $\Delta \text{y}$ and $\Delta \text{z}$ represent the quark spin polarizations due to valence quarks.
The expressions for orbital moments of $u,d$ and $s$, i.e., $\mu\left(u^+\rightarrow \right)\ $, $\mu\left(d^+\rightarrow \right)\ $ and $\mu\left(s^+\rightarrow \right)\ $ in terms of effective masses of quarks (in units of nuclear magneton $\mu_N$) are given as
\begin{eqnarray}
\mu\left(u^+\rightarrow \right)=&&a\left[\frac{-m_{\pi}^2+3m_{u}^{*2}}{2m_{\pi}\left(m_u^*+m_{\pi}\right)} 
- \frac{\varepsilon^{2}(m_{K}^2-3m_{u}^{*2})}{2m_{K}\left(m_u^*+m_{K}\right)}\right] \nonumber \\
&&+a\left[\frac{(3+\varpi^{2}+2\tau^{2})m_{\eta^{'}}^{2}}{6m_{\eta^{'}}\left(m_u^*+m_{\eta^{'}}\right)}\right],
\end{eqnarray}
\begin{eqnarray}
\mu\left(d^+\rightarrow \right)=&&a\frac{m_u^*}{m_d^*}\left[\frac{2m_{\pi}^2-3m_{d}^{*2}}{2m_{\pi}\left(m_d^*+m_{\pi}\right)}-\frac{\varepsilon^{2}m_{K}^2}{2m_{K}\left(m_d^*+m_{K}\right)}\right] \nonumber \\
&&-a\frac{m_u^*}{m_d^*}\left[\frac{(3+\varpi^{2}+2\tau^{2})m_{\eta^{'}}^{2}}{12m_{\eta^{'}}\left(m_u^*+m_{\eta{'}}\right)}\right],
\end{eqnarray}
\begin{eqnarray}
\mu\left(s^+\rightarrow \right)=a\frac{m_u^*}{m_s^*}\left[\frac{\varepsilon^{2}(m_{K}^2-3m_s^{*2})}{2m_{K}\left(m_s^*+m_{K}\right)}
-\frac{(2\varpi^{2}+\tau^{2})m_{\eta^{'}}^{2}}{6m_{\eta^{'}}\left(m_u^*+m_{\eta{'}}\right)}\right]. \nonumber \\
\end{eqnarray}  
 
 These contributions can be calculated as done in Ref. \cite{har}. However, it is worth noting that in order to consider the medium modification of sea quark spin polarization $\Delta q_{\rm sea}$ and orbital angular momentum contributions  $\mu\left(u^+\rightarrow \right)$, $\mu\left(d^+\rightarrow \right)$ and $\mu\left(s^+\rightarrow \right)$. The parameters $\varepsilon$ and $\varpi$ appear in the linear representation of octet scalar density \cite{ling}. The linear combination of these parameters  give the familiar `F' and `D' coefficients. 
These parameters can be expressed in terms of medium modified quark and baryon masses as 
\begin{equation}
\varepsilon=\frac{M_{\Sigma}^{*}-M_{\Xi}^{*}}{\left(\frac{m_{u}^*+m_{d}^*-2m_{s}^*}{2}\right)}, \label{alphadash}
\end{equation} 
and 
\begin{equation}
\varpi=\frac{M_{\Sigma}^{*}-M_{N}^{*}}{\left(\frac{m_{u}^*+m_{d}^*-2m_{s}^*}{2}\right)}. \label{betadash}
\end{equation} 
These two parameters along with $\tau$ given by equation (\ref{zetadash}) lead to medium modification of sea quark polarizations and orbital moments. Physically $\varepsilon^2a$, $\varpi^2a$ and $\tau^2a$ respectively denote the probabilities of transitions $u(d)\rightarrow s+K^-$, $u(d,s)\rightarrow u(d,s)+\eta$ and $u(d,s)\rightarrow u(d,s)+\eta^{'}$. Note that orbital angular momentum contribution is calculated using the parameters $\varepsilon,\varpi$ and $\tau$ along with masses of GBs. The GBs contributions are dominated by pion contribution as compared to contributions from other GBs.   
\section{Numerical results} \label{results}
In this section we present the results of our investigation on magnetic moment of baryons at finite density and temperature of medium. Various parameters used in the present work are tabulated in table (\ref{cc}).
\begin{table}[h]
\resizebox{.5\hsize}{!}{ 
\begin{tabular}{|c|c|c|c|c|}
\hline 
$m_u$ (MeV) & $m_d$ (MeV) & $m_s$ (MeV) & $m_{\pi}$ (MeV) & $m_K$ (MeV) \\ 
\hline 
313 & 313 & 490 & 139 & 494  \\ 
\hline
\hline 
$k_0$ & $k_1$ & $k_2$ & $k_3$ & $k_4$  \\ 
\hline 
4.94 & 2.12 & -10.16 & -5.38 & -0.06  \\ 
\hline
\hline 
$\sigma_0$ (MeV) & $\zeta_0$ (MeV) & $\chi_0$ (MeV) & $\xi$ & $\rho_0$ ($\text{fm}^{-3}$)  \\ 
\hline 
-92.8 & -96.5 & 254.6 & 6/33 & 0.16  \\ 
\hline
\hline 
$g_{\sigma}^u=g_{\sigma}^d$ & $g_{\sigma}^s$ & $g_{\zeta}^u=g_{\zeta}^d$ & $g_{\zeta}^s$ & $g_4$\\ 
\hline 
3.37 & 0 & 0 & 4.77 & 37.4 \\ 
\hline
\end{tabular}}
\caption{Values of various parameters used in the present work \cite{wang}.} \label{cc}
\end{table} 

\begin{figure}[h]
\resizebox{0.75\textwidth}{!}{
\includegraphics{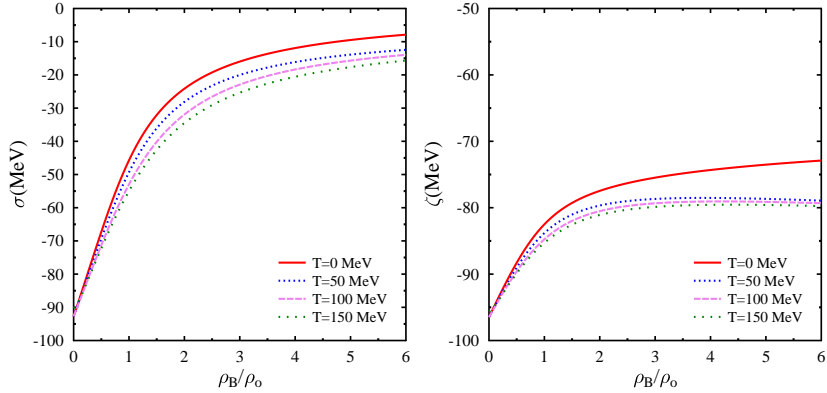}
}
\caption{{$\sigma$ and $\zeta$ fields (at T=0, 50, 100 and 150 MeV) versus baryonic density (in units of nuclear saturation density $\rho_0$).} } \label{fields}
\end{figure} 
\begin{figure}
\resizebox{0.8\textwidth}{!}{
\includegraphics{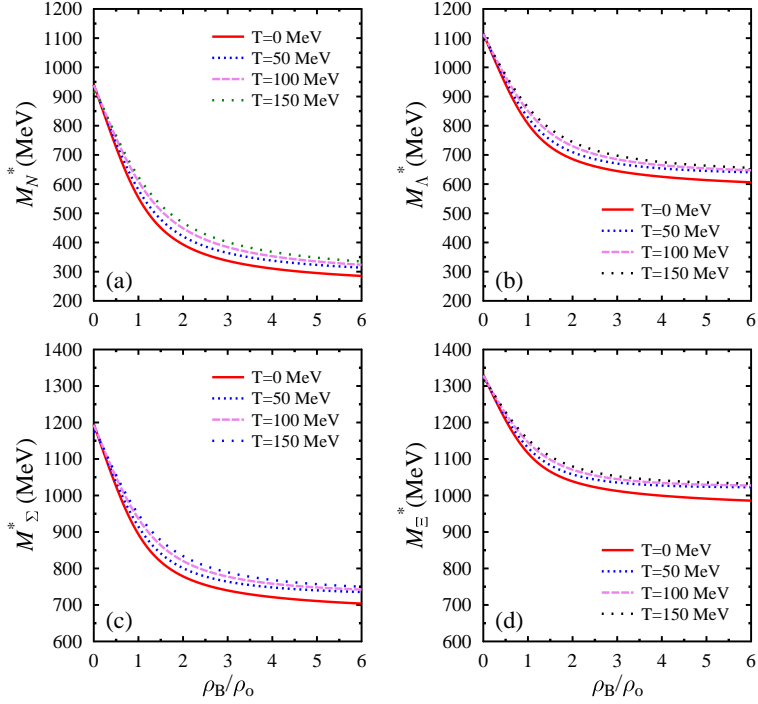} }
\caption{{Effective masses of octet baryons (at T=0, 50, 100 and 150 MeV) versus baryonic density (in units of nuclear saturation density $\rho_0$).} } \label{bmasst}
\end{figure}
\begin{figure}
\resizebox{0.75\textwidth}{!}{
\includegraphics{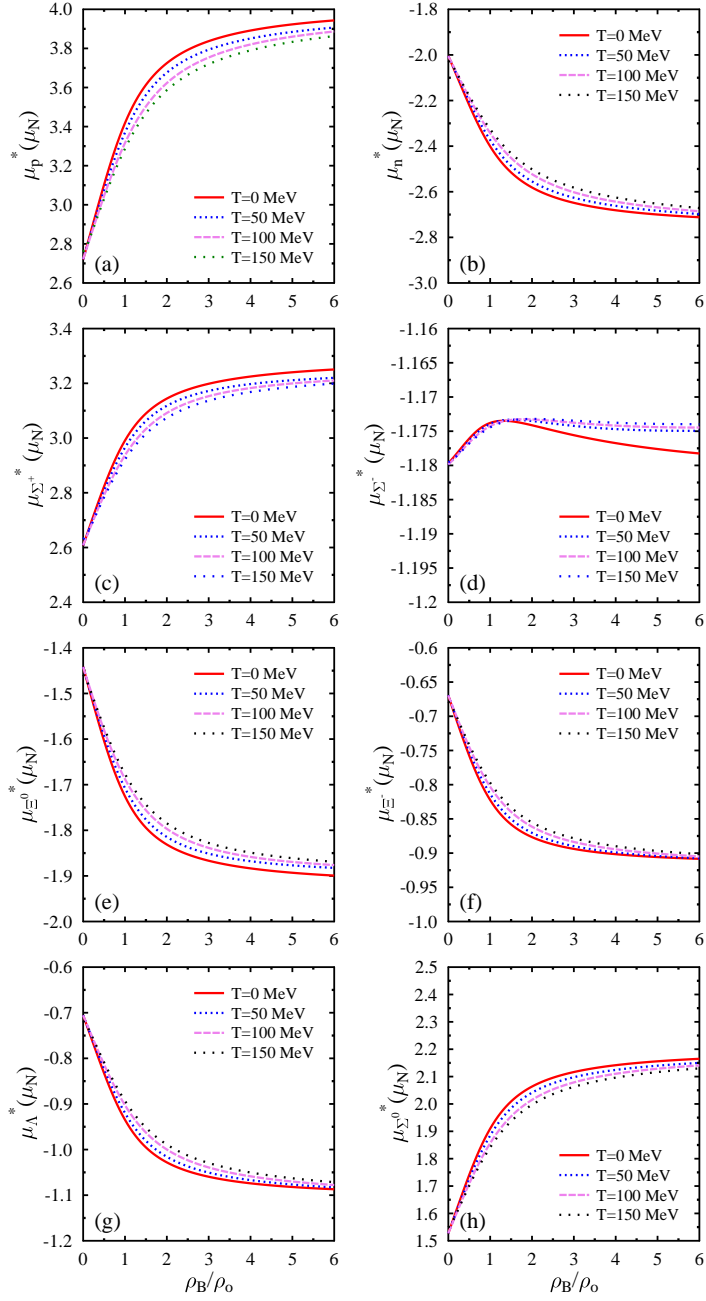} }
\caption{{Magnetic moment of octet baryons (at T = 0, 50, 100 and 150 MeV) versus baryonic density (in units of nuclear saturation density $\rho_0$).} } \label{mmt}
\end{figure}
\begin{figure}
\resizebox{0.75\textwidth}{!}{
\includegraphics{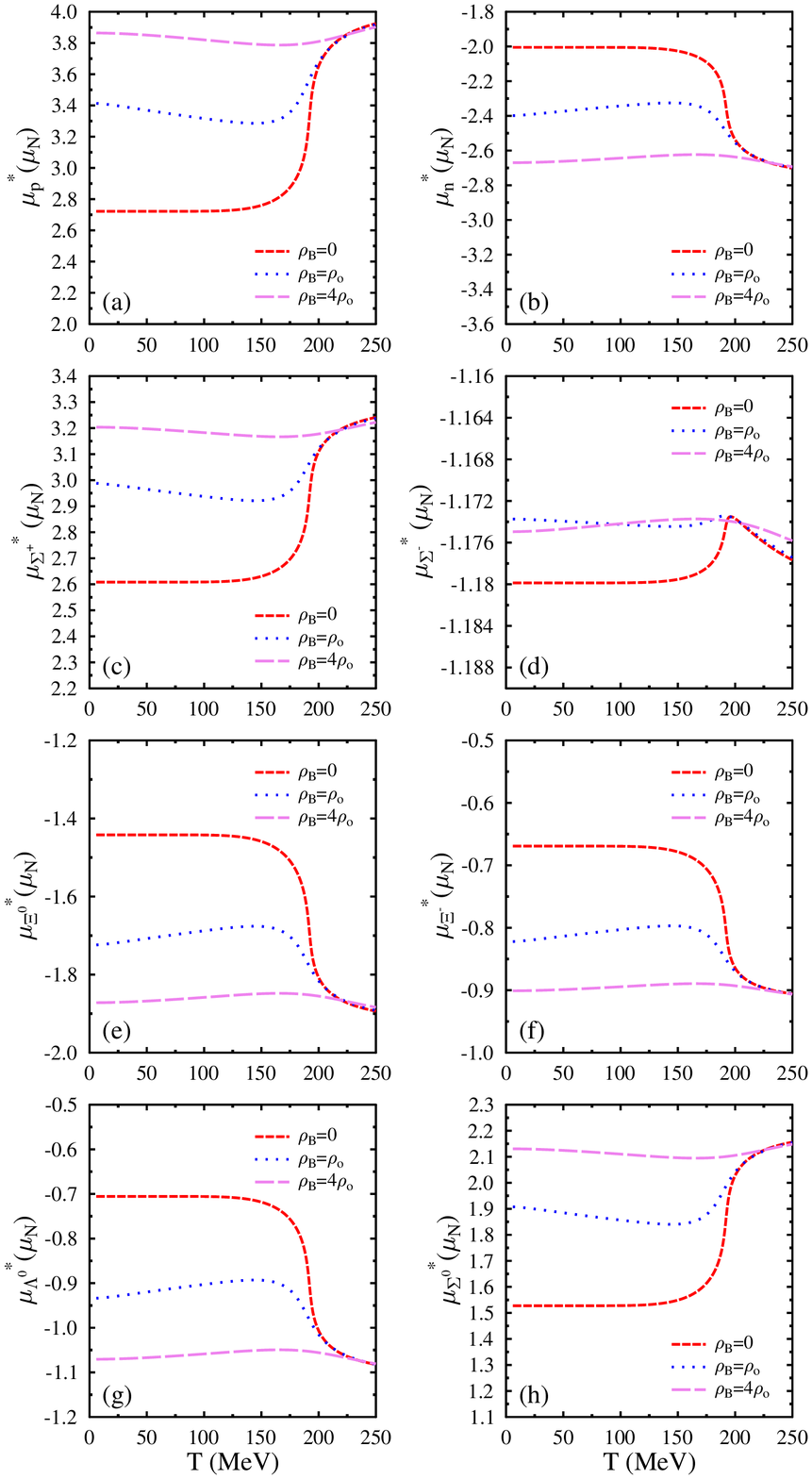} }
\caption{{Magnetic moments of baryons as a function of temperature at $\rho_B=0,\rho_0$ and $4\rho_0$.} } \label{mmpt}
\end{figure}
 
From equation (\ref{magandmass}) it is clear that the value of magnetic moment of constituent quarks depends on the effective masses of the quarks and baryons, which in-turn depends on the scalar fields $\sigma$ and $\zeta$ through equations (\ref{qmass}), (\ref{qenergy}), (\ref{baryonmass}) and (\ref{energy}). In order to study the effect of density, on the scalar fields $\sigma$ and $\zeta$, in the left panel of fig. (\ref{fields}), we plot the scalar field $\sigma$ with nuclear matter density $\rho_B$ (in the units of nuclear saturation density), at different temperatures of the medium T = $0, 50, 100$ and $150$ MeV. We observe that the magnitude of $\sigma$ field decreases sharply with the rise of nuclear matter density from upto $\rho_B=2\rho_0$. For densities more than $2\rho_0$, the decrease in magnitude of $\sigma$ field as a function of $\rho_B$ is slow. For example, at T = 0 MeV, the $\sigma$ field changes by $73\%$ as $\rho_B$ is changed from zero to $2\rho_0$. However, as the baryonic density increases from $2\rho_0$ to $4\rho_0$, the magnitude of scalar field $\sigma$  changes by $50\%$. The amount of this decrease in the value of $\sigma$ field is even lesser at higher densities. Considering the effect of temperature, we observe that the magnitude of $\sigma$-field decreases less rapidly, as a function of density at higher temperatures of medium as compared to lower temperatures. 
However, at $\rho_B=0$, with the rise of temperature the magnitudes of scalar fields decrease. At given finite density of the medium the magnitude of $\sigma$ field increases with the rise of temperature. For example, at nuclear saturation density ($\rho_B=\rho_0$), magnitudes of $\sigma$ fields are $45.71,49.3$ and $52.96 $ MeV at T = 0, 50 and 100 MeV, respectively. This is explained as follows. At $\rho_B=0$, the thermal distribution functions alone effect the variation of fields. However, with the rise of density, another contribution starts coming from higher momentum states, which provides opposite effect to the variation of scalar fields  \cite{AMU}. Thus, due to these two contributions, i.e., thermal distribution function and higher momentum states, the behavior of scalar fields is reversed with the rise of temperature, at finite value of density of medium as compared to its behavior at zero baryonic density.
  
In the right panel of fig. (\ref{fields}), we have plotted the variation of $\zeta$ field with nuclear matter density at temperatures T = $0, 50, 100$ and $150$ MeV. One can clearly see that the magnitude of $\zeta$ field decreases very slowly as a function of density as compared to scalar field $\sigma$ indicating that there is strong correlation between the nucleons and the $\sigma$ field. However, the $\zeta$ field changes very slowly because of absence of its dependence on non-strange quark content of the medium. For example, at T = 0 MeV, there is decrease of only $20\%$ in magnitude of $\zeta$ field as $\rho_B$ increases from 0 to $2\rho_0$. Further, on calculating the variation in the magnitude of $\zeta$ field at different temperatures, one can note that decrease in magnitude as a function of density is less for higher temperatures as compared to T = 0 MeV. This difference increases for higher values of nuclear matter density. For example, at $\rho_B=\rho_0$, there is difference of $2.25$ MeV in magnitudes of $\zeta$ field at T = 0 MeV and T = 100 MeV. However, at $\rho_B=5\rho_0$, this difference in the values of $\zeta$ field changes to $\text{5.6}$ MeV. At nuclear saturation density, the magnitude of $\zeta$ field decreases by only $2\%$ with the rise of temperature from T = 0 MeV to T = 100 MeV. However, at $\rho_B=5\rho_0$ the above value of percentage change shift to $6\%$. 

Using the above calculated values of $\sigma$ and $\zeta$ fields, the in-medium quark masses, $m_q^*$, can be evaluated using equation (\ref{qmass}). Note that in this work, the non-strange quark masses ($m_u^*$ and $m_d^*$) depend on scalar meson field $\sigma$ only. As the coupling constant $g_{\zeta}^u=g_{\zeta}^d=0$, therefore in equation (\ref{qmass}), $\zeta$ is eliminated for $m_{u}^*$ and $m_{d}^*$. As magnitude of $\sigma$ field decreases sharply with the rise of density especially at densities upto $2\rho_0$, there is steep decrease in the effective mass of non-strange quarks at a lower value of density medium for a fixed value of temperature. Whereas at higher nuclear matter density, the decrease in effective quark mass is quite less. For example, at temperature T = 0 MeV, effective mass of `$u$'(or `$d$') quark at $\rho_B=3\rho_0, 4\rho_0$ and $5\rho_0$  decreases to $\text{53.92}$, $\text{40.19}$ and $\text{32.01}$ MeV, respectively from its vacuum value of $\text{313}$ MeV. 

For a given finite value of baryonic density, the effective mass of non-strange quarks increases with the increase in the temperature of nuclear medium. For example, at $\rho_B=\rho_0$, the values of $m_u^*$ are observed to be 154.2, 165.9, 178.2 and 185 MeV at T = 0, 50, 100 and 150 MeV, respectively.  
However, at zero density, the effective masses of non-strange quarks decrease with the rise of temperature. The reason for this behavior is the dynamical generation of  masses of quarks by coupling with scalar fields $\sigma$ and $\zeta$.  

 Further, the magnitudes of scalar fields decrease with the rise of density, at T = 0 MeV. Therefore, the value of $m_u^*$ (and $m_d^*$) also decreases with rise of baryonic density, at T = 0 MeV. The probable cause behind this behavior of effective quark masses can be the chiral symmetry restoration at higher densities, which has been reported in literature, by using chiral hadronic model, in the quark degrees of freedom \cite{rau}. 
 
It has been seen that $m_s^*$ decreases less rapidly as compared to $m_u^*$ and $m_d^*$, when  plotted as a function of baryonic density, at given value of temperature. At temperature T = 0 MeV, as the  density of medium increases from 0 upto $\rho_0$, $m_s^*$ decreases by about $14\%$. Further, at more higher values of density but at same temperature (T = 0 MeV), $m_s^*$ decreases very slowly. The reason for this behavior of $m_s^*$ at finite baryonic density is its dependence on scalar $\zeta$ field, and, the absence of coupling between $s$-quark and $\sigma$ field as $g_{\sigma}^s=0$.

One also finds that the effective mass of `$s$' quark increases with the rise of temperature, at given finite value of density. For example, at $\rho_B=\rho_0$, the effective masses of `$s$' quark are $\text{422.5}$, $\text{427.5}$ and $\text{432.5}$ MeV at temperatures T = $0, 50$ and $100$ MeV, respectively. Further, for higher values of density, the increase in effective mass of `$s$' quark becomes slow with rise of temperature. For example, at $\rho_B=2\rho_0$, for the rise of temperature from T = 0 to T = 50 MeV and from T = 50 to T = 100 MeV, the effective mass of `$s$' quark increases by $\text{9.7}$ MeV and $\text{4}$ MeV, respectively.

Also, we observe that the effective mass of `$s$' quark decreases with rise of density upto baryonic density $\rho_B=4\rho_0$, at finite temperature. However, on further increase of density  above $4\rho_0$, at the same value of finite temperature, the effective mass of `$s$' quark starts increasing. The increase in constituent quark masses with increase in density above $4\rho_0$ of medium, at given finite temperature can be due to deconfinement phase transition at higher density \cite{abu}.

Now we discuss the medium modification of octet baryon masses calculated using equation (\ref{baryonmass}), 
through medium modified masses of quarks in equation (\ref{qmass}). In fig. (\ref{bmasst}), we plot the medium modified octet baryon masses as ($M_i^*$, $i=N, \Sigma, \Xi , \Lambda$) a function of density at temperatures, T = $0, 50, 100$ and $150$ MeV. We see that the variation of effective masses of constitute quarks largely effects the medium modification of baryonic masses. For example, in case of nucleons having non-strange quark content only, there is steep decrease in the effective baryonic masses. Our calculations show that at T = 0 MeV, as the density increases from $\rho_B=0$ to $\rho_0$, the effective mass of nucleons decreases by $41\%$ from its vacuum value. The similar behavior has been reported in literature \cite{effective}, where the effective field calculations show the decrease of $30\%$ in effective nucleon masses for the rise of density from  $\rho_B=0$ to $\rho_0$, at T = 0 MeV. This difference is due to model dependence of quark and baryon masses. Further, at T = 0 MeV, the effective nucleon mass decreases to $41\%$, $36\%$, $33\%$ and $31\%$ of its vacuum value at $\rho_B=2\rho_0, 3\rho_0, 4\rho_0$ and $5\rho_0$, respectively. 

As compared to nucleons, the in-medium masses of strange baryons decrease less rapidly as a function of density of medium, at given temperature. For example, at T = 0 MeV, for rise of density from $\rho_B=0$ to $\rho_0$, there is decrease of $25\%$, $28\%$ and $16\%$ in the effective masses of $\Sigma$, $\Lambda$ and $\Xi$ baryons, respectively.  

One can also observe that, with the rise of temperature, at given value of density of medium, the effective masses of baryons increase. For example, at $\rho_B=\rho_0$, effective mass of nucleons increase by $10\%$ as the temperature rises from T = 0 to T = 100 MeV.  
Further, at $\rho_B=\rho_0$, the effective masses of $\Sigma$, $\Lambda$ and $\Xi$ baryons increase by $5\%$, $5.5\%$ and $3\%$, respectively, with the rise of temperature from T = 0 MeV to T = 100 MeV. One can see that the increase of effective masses of strange baryon masses with the rise of temperature, at given finite value of density of medium, is slow as compared to increase in effective masses of nucleons. The reason behind this behavior of effective masses of octet baryons is their dependence on the constituent quark masses. The effective masses of `$u$' and `$d$' quarks increase significantly, whereas effective mass of `$s$' quark increases slowly, with the rise of temperature, at finite value of density. This is why the increase in effective masses of strange baryons  slows with increase in strangeness content of the baryon.     

Now we will discuss the medium modification of baryon magnetic moments of octet baryons. In figure (\ref{mmt}), we plot magnetic moment of octet baryons with density at temperatures T = $0, 50, 100$ and $150$ MeV. In tables (\ref{mmval}) and (\ref{mmvaldifft}), we have given the observed values of medium modified magnetic moments of octet baryons, at temperatures T = 0 MeV and 100 MeV, respectively. The values are calculated for densities $\rho_B=0, \rho_0$ and $4\rho_0$. Note that in table (\ref{mmval}) we have also given the experimentally observed vacuum values of magnetic moment of octet baryons. 

If we consider the effect of valence quarks only, the vacuum value of magnetic moment of baryons as calculated in our model comes out to be larger than the experimental values. For example, at $\rho_B=0$ and T = 0 MeV, considering the valence quark effect only, the magnetic moment of proton comes out to be $2.994\mu_N$, which is more than the experimental value of magnetic moment of proton in vacuum, i.e., $2.79\mu_N$ \cite{har}. In order to get the more realistic values of magnetic moments, we have included the contribution from the `Goldstone Boson Exchange' effect, also known as sea quark effect, whose contribution to the magnetic moment of baryons is opposite to that of the valence quark contribution. Following Cheng and Li mechanism \cite{cheng}, we have also considered the effect of the contribution of the orbital angular momentum of sea quarks \cite{har}. It is important to note that sea quark effect gives opposite contribution to total magnetic moment of baryons as compared to valence quark effect, whereas, the contribution from orbital angular momentum of sea quarks is of the same sign as that from the valence quark effect.

 The observed behavior of magnetic moment of baryons may be directly related to the spin decomposition of nucleons and other baryons, 
which is one of the key problems in nucleon structure physics \cite{buch,ref1spin,ref2spin}. The spin sum rule to calculate proton spin J can be expressed as 
 \begin{equation}
 J=\frac{1}{2} \Sigma+L_{q}+\Delta g+L_{g}, \nonumber
 \end{equation}
where `$\Sigma$' is the quark spin, `$L_q$' is quark angular momentum, `$\Delta g$' is the contribution from gluon spin and `$L_{g}$' is orbital angular momentum of gluon. Experimental observations by European Muon Collaboration in deep inelastic scattering experiments have shown that valence quarks in proton carry only about $30\%$ of total spin of proton \cite{rana}. The remaining spin may come from angular momentum part of quark spin, gluon spin part and orbital angular momentum of gluon in the total spin of proton. The quark spin ($\Sigma$) may further split into the contribution from valence and sea quarks as $\Sigma=\Sigma_V+\Sigma_S$. Gluon spin and orbital angular momentum of gluon parts are very small as indicated by different experimental studies \cite{gluon1,gluon2}, and can be neglected at present. 
  In the present model, the splitting of quark into quark and GB leads to the flip of quark spin which means that the quarks produced through this process which constitute `quark sea' are eventually polarized in the opposite direction to that of the valence quarks. 
The contribution from orbital angular momentum part is however of the same sign as that of the valence quarks.  
 Further, in case of proton due to flavor asymmetry, the effect of polarization of two `u' quarks is more than the effect of polarization of one `d' quark. This leads to the fact that in case of proton total contribution from sea quark polarization is more than the opposite contribution from orbital angular momentum part. This behavior for the spin sum rule has been reported in literature \cite{cheng1995,lingfong,har3,song1}, and the magnetic moments calculated in the present work also follows the same behavior.  
 
On comparing the values in tables (\ref{mmval}) and (\ref{mmvaldifft}), we find that at $\rho_B=0$, the magnetic moments of baryons are almost same at T = 0 MeV and T = 100 MeV. This means that at zero baryonic density there is negligible effect of rise of temperature on effective magnetic moments of baryons. However, at finite densities, there is a noticeable change in the values of magnetic moments of baryons specially in case of nucleons. 

We find that with the rise of density of medium at T = 0 MeV as well as 100 MeV, the magnitude of sea quark polarizations decrease. For example, at T = 0 MeV, the magnitude of $\Delta u_{\rm sea}$ in case of proton as given in equation (\ref{seapol}) is found to be $0.165,0.134$ and $0.129$ at $\rho_B=0,\rho_0$ and $4\rho_0$, respectively. This is due to medium modification of symmetry breaking parameters $\varepsilon$ and $\varpi$ along with parameter $\tau$. However, if we do not consider medium modification of these parameters, the value of $\Delta u_{\rm sea}$ remains equal to its vacuum value at all the densities. However, with the rise of density of medium at same temperature, the total effective magnetic moments vary significantly because of medium modification of sea quark polarization. For example, at T = 0 MeV and $\rho_B=\rho_0$, $\mu_{p}=3.418\mu_N$ and $\mu_{\Xi^0}=-1.726\mu_N$ with medium modified sea quark polarization, whereas, with constant value(vacuum value) of sea quark polarization at all densities, these values comes out to be $3.450$ and $-1.902\mu_N$, respectively.
If we consider the effect of rise of temperature on $\Delta u_{\rm sea}$, we find that for given finite value of density of medium $\Delta u_{\rm sea}$ is negligibly affected by the rise of temperature. For example, at T = 100 MeV, magnitude of $\Delta u_{\rm sea}$ are $0.165,0.137$ and $0.128$ at $\rho_B=0,\rho_0$ and $4\rho_0$, respectively. 
Further, one can see that both at T = 0 MeV and 100 MeV, the contribution of orbital angular momentum of sea quarks decreases with the rise of density.    
 
 \begin{table*}[ht]
 \resizebox{.95\hsize}{!}{ 
 \begin{tabular}{|c|c|c|c|c|c|c|c|c|c|c|c|c|c|}  
 \hline 
 & Data \cite{experiment} & \multicolumn{4}{c|}{$\rho_B=0$} & \multicolumn{4}{c|}{$\rho_B=\rho_0$} & \multicolumn{4}{c|}{$\rho_B=4\rho_0$}   \\ 
 \hline
 & $\mu_{\rm total}$ & $\mu_{\text{val}}$ & $\mu_{\text{sea}}$ & $\mu_{\text{orbital}}$ & $\mu_{\text{total}}$ & $\mu_{\text{val}}$ & $\mu_{\text{sea}}$ & $\mu_{\text{orbital}}$ & $\mu_{\text{total}}$ & $\mu_{\text{val}}$ & $\mu_{\text{sea}}$ & $\mu_{\text{orbital}}$ & $\mu_{\text{total}}$  \\ 
 \hline 
 $\mu_p^*(\mu_N)$ & $2.792$ & $2.994$ & $-0.724$ & $0.450$ & $2.72$ & $4.232$ & $-0.985$ & $0.172$ & $3.418$ & $5.008$ & $-1.129$ & $0.013$ & $3.892$   \\ 
 \hline 
 $\mu_n^*(\mu_N)$ & $-1.913$ &$-1.996$ & $0.414$ & $-0.422$ & $-2.004$ & $-2.821$ & $0.583$ & $-0.166$ & $-2.401$ & $-3.338$ & $0.656$ & $0.001$ & $-2.681$   \\ 
 \hline 
 $\mu_{\Sigma^+}^*(\mu_N)$ & $2.458$ &$3.002$ & $-0.776$ & $0.381$ & $2.607$ & $3.758$ & $-0.914$ & $0.146$ & $2.991$ & $4.190$ & $-0.980$ & $0.015$ & $3.225$  \\ 
 \hline 
 $\mu_{\Sigma^-}^*(\mu_N)$ & $-1.160$ &$-1.001$ & $0.137$ & $-0.316$ & $-1.18$ & $-1.253$ & $0.201$ & $-0.122$ & $-1.174$ & $-1.397$ & $0.215$ & $0.005$ & $-1.177$   \\ 
 \hline 
 $\mu_{\Sigma^0}^*(\mu_N)$ & $-1.610$ &$-1.386$ & $-0.078$ & $-0.256$ & $-1.721$ & $-1.909$ & $-0.029$ & $-0.096$ & $-2.035$ & $-2.312$ & $-0.016$ & $-0.003$ & $-2.331$   \\ 
 \hline 
 $\mu_{\Xi^0}^*(\mu_N)$ & $-1.250$ & $-2.000$ & $0.614$ & $-0.055$ & $-1.441$ & $-2.323$ & $0.623$ & $-0.025$ & $-1.726$ & $-2.498$ & $0.621$ & $-0.006$ & $-1.883$   \\ 
 \hline 
 $\mu_{\Xi^-}^*(\mu_N)$ & $-0.650$ & $-1.000$ & $0.386$ & $-0.055$ & $-0.669$ & $-1.162$ & $0.365$ & $-0.025$ & $-0.822$ & $-1.249$ & $0.354$ & $-0.006$ & $-0.902$   \\ 
 \hline 
 $\mu_{\Lambda}^*(\mu_N)$ & $-0.613$ & $-0.664$ & $0.336$ & $-0.042$ & $-0.705$ & $-0.952$ & $0.362$ & $-0.019$ & $-0.935$ & $-1.070$ & $0.370$ & $-0.004$ & $-1.074$   \\ 
 \hline 
 \end{tabular}}
\caption{Effective magnetic moments of octet baryons at T = 0 MeV and $\rho_B=0, \rho_0$ and $4\rho_0$.} \label{mmval}
\end{table*} 

We also see that with the increase of strangeness content the increase in magnitude of effective magnetic moment of baryon is less. This is because $m_s^*$ varies very slowly with density at given temperature.   
Further, at given finite temperature, the effective magnetic moments are not much sensitive to quark mass variation for higher densities. 
Our calculations show that at temperature T = 0 MeV, for the rise of density of nuclear medium from $\rho_B=0$ to $\rho_0$, effective magnetic moment of proton increases by $26\%$. However, for further increase in density of medium, at the same temperature, the rise of magnetic moment of proton becomes slow. For example, at T = 0 MeV, for rise of density from $2\rho_0$ to $6\rho_0$ 
the effective magnetic moment rise by $20\%$. A cloudy bag model prediction shows enhancement of magnetic moment with the rise of nuclear matter density from $\rho_B=0$ to $\rho_0$ in the range of $2-20\%$ \cite{ryu3}. Further, the models like constituent quark model, QMC model pion cloud, skyrme model, chiral quark soliton model and NJL model predict enhancement upto $10\%$ . In our calculations this enhancement is $26\%$, which is quite large as compared to the previous predictions. This is due to model dependence of in effective baryon masses and hence magnetic moments.

Further, in table (\ref{mmval}), we see that at T = 0 MeV, for rise of density from $\rho_B=0$ to nuclear saturation density, the magnitude of effective magnetic moments increases by $15\%$, $0.5\%$ and $25\%$ in case of $\Sigma^{+}$, $\Sigma^{-}$ and $\Sigma^{0}$ baryons respectively. The very small change in effective magnetic moment of $\Sigma^{-}$ is due to comparable contributions from sea quark effect and orbital angular momentum of sea quarks, whereas in in case of other baryons these contributions do not completely cancel out each other. In case of $\Xi^{0}$, $\Xi^{-}$  baryons, this increase in magnitude of magnetic moments is $20\%$ and $23\%$, respectively. In particular for $\Lambda$ baryon the magnitude of effective magnetic moment increases by $32\%$. This behavior is completely different from that in case of QMC calculations, where the magnitude of $\mu_{\Lambda}^*$ decreases by $0.7\%$. However, in case of modified QMC calculations the magnitude increases by $10\%$ \cite{ryu3}. The possible reason for this can be model dependence of effective quark masses.
In the present work, the modification of magnetic moments of baryons depend on medium modification of constituent quark masses, whereas in Ref. \cite{ryu3}, the modification of magnetic moments were derived from modification of bag radius. 

 \begin{table*}[hb]
  \begin{tabular}{|c|c|c|c|c|c|c|c|c|c|c|c|c|}  
 \hline 
 &  \multicolumn{4}{c|}{$\rho_B=0$} & \multicolumn{4}{c|}{$\rho_B=\rho_0$} & \multicolumn{4}{c|}{$\rho_B=4\rho_0$}   \\ 
 \hline
 & $\mu_{\text{val}}$ & $\mu_{\text{sea}}$ & $\mu_{\text{orbital}}$ & $\mu_{\text{total}}$ & $\mu_{\text{val}}$ & $\mu_{\text{sea}}$ & $\mu_{\text{orbital}}$ & $\mu_{\text{total}}$ & $\mu_{\text{val}}$ & $\mu_{\text{sea}}$ & $\mu_{\text{orbital}}$ & $\mu_{\text{total}}$  \\ 
 \hline 
 $\mu_p^*(\mu_N)$ & $3.001$ & $-0.727$ & $0.449$ & $2.723$ & $4.053$ & $-0.948$ & $0.212$ & $3.316$ & $4.873$ & $-1.090$ & $0.037$ & $3.820$   \\ 
 \hline 
 $\mu_n^*(\mu_N)$ & $-2.001$ & $0.416$ & $-0.421$ & $-2.005$ & $-2.702$ & $0.560$ & $-0.201$ & $-2.343$ & $-3.248$ & $0.631$ & $-0.026$ & $-2.643$   \\ 
 \hline 
 $\mu_{\Sigma^+}^*(\mu_N)$ & $3.007$ & $-0.778$ & $0.380$ & $2.609$ & $3.651$ & $-0.894$ & $0.180$ & $2.937$ & $4.098$ & $-0.950$ & $0.035$ & $3.183$  \\ 
 \hline 
 $\mu_{\Sigma^-}^*(\mu_N)$ & $-1.002$ & $0.138$ & $-0.316$ & $-1.18$ & $-1.217$ & $0.193$ & $-0.150$ & $-1.174$ & $-1.366$ & $0.207$ & $-0.015$ & $-1.174$   \\ 
 \hline 
 $\mu_{\Sigma^0}^*(\mu_N)$ & $-1.389$ & $-0.080$ & $-0.255$ & $-1.724$ & $-1.826$ & $-0.033$ & $-0.119$ & $-1.979$ & $-2.228$ & $0.001$ & $-0.018$ & $-2.244$   \\ 
 \hline 
 $\mu_{\Xi^0}^*(\mu_N)$ & $-2.004$ & $0.625$ & $-0.055$ & $-1.442$ & $-2.278$ & $0.620$ & $-0.029$ & $-1.688$ & $-2.446$ & $0.596$ & $-0.009$ & $-1.858$   \\ 
 \hline 
 $\mu_{\Xi^-}^*(\mu_N)$ & $-1.002$ & $0.388$ & $-0.055$ & $-1.228$ & $-1.139$ & $0.366$ & $-0.029$ & $-1.546$ & $-1.223$ & $0.337$ & $-0.009$ & $-1.771$   \\ 
 \hline 
 $\mu_{\Lambda}^*(\mu_N)$ & $-1.002$ & $0.338$ & $-0.042$ & $-0.706$ & $-1.238$ & $0.357$ & $-0.022$ & $-0.903$ & $-1.405$ & $0.352$ & $-0.007$ & $-1.059$   \\ 
 \hline 
 \end{tabular}
\caption{Effective magnetic moments of octet baryons at T = 100 MeV and $\rho_B=0, \rho_0$ and $4\rho_0$.} \label{mmvaldifft}
\end{table*} 
%
%
 

To understand more explicitly the effect of temperature of the medium on magnetic moments of octet baryons, in fig. (\ref{mmpt}), we plot the effective magnetic moments of baryons as a function of temperature, at $\rho_B=0$, $\rho_0$ and $4\rho_0$. 

We note that at given density of medium, with the rise of temperature, the magnetic moments of baryons increase slightly. For example, at $\rho_B=0$, effective values of magnetic moment of proton are observed to be  $2.720\mu_N, 2.722\mu_N, 2.723\mu_N, 2.760\mu_N$ at temperatures, T = $0, 50, 100$ and $150$ MeV, respectively. Hence, the variation in effective magnetic moment of baryons as a function of temperature is negligible at zero density upto critical temperature. These results are in good agreement with those obtained in Ref. \cite{christov,abu}, where magnetic moment of nucleons were calculated using quark sigma model. However, as the temperature reaches its critical value there is steep increase in magnitude of effective magnetic moments. This can be attributed the second order phase transition above critical temperature. 
     
At finite density, the change in effective value of magnetic moment of baryons is almost negligible as a function of temperature as compared to that at zero density. These results can be explained as follows. From equations (\ref{mag}) and (\ref{magandmass}), we find that the effective magnetic moment of baryons are inversely proportional to the medium modified values of constituent quark masses. At $\rho_B=0$, the effective quark mass remains almost same with the rise of temperature upto certain value of temperature, because the thermal distribution functions alone effect the self energies of constituent quarks and hence decreasing the effective quark masses (increasing the effective magnetic moment of baryons). However, with the rise of density, another contribution starts coming from higher momentum states due to which the effective magnetic moments start decreasing (as the effective masses of quarks increase) \cite{AMU}. Further, for still high densities, i.e., $4\rho_0$ or more, the variation of effective magnetic moment of baryons become insensitive to the variation in effective mass of constituent quarks. This can be due to second order phase transition at higher densities and temperatures. This observation is further justified by those expected in Ref. \cite{ryu1}, where medium modified baryonic magnetic moments using modified quark meson coupling model were calculated.

\section{Summary} \label{summ}
We have studied the magnetic moment of baryons at finite density and temperature of symmetric nuclear matter by using chiral SU(3) quark mean field approach. The 
explicit contributions from valence quarks, sea quarks and orbital angular momentum of sea quarks have also been considered to give better insight into medium modification of magnetic moments. The consideration of valence quark effect only, gives magnetic moments more than the experimental data for vacuum values. The sea quark effect gives opposite contribution to the total effective magnetic moments, as compared to that by valence quarks. However, considering the sea quark effect alone decreases the vacuum values lower than those in experimental data \cite{har}. Hence, in order to get more realistic vacuum values we have considered the contribution from orbital angular momentum of sea quarks, which gives considerable opposite contribution to magnetic moments as compared to that from sea quarks especially at lower densities and small contribution at higher densities.     

 Magnetic moment of nucleons are found to vary largely as a function of density at low temperatures, however, at higher temperature this variation of magnetic moment becomes slow. The magnetic moments of strange baryons are found to vary slowly with density as well as temperature as compared to those of non-strange baryons. The reason behind this behavior of magnetic moments is their dependence on medium modified  values of strange quark mass, which vary very slowly because of small coupling  with the scalar meson field. Further, the variation of effective magnetic moments of baryons as a function of temperature is negligible for nuclear matter density higher than $4\rho_0$. This indicates second order phase transition at higher densities \cite{wang neu}. 
 
It is found in Ref. \cite{loop} that the pion loop correction shows only a minute contribution to anomalous magnetic moments of baryons. However, we have derived the medium modification of sea quark polarization through medium modification of symmetry breaking parameters $\varepsilon$ and $\varpi$. The results can be further improved by including contribution from effects from relativistic and exchange currents \cite{mg}, pion cloud contributions \cite{thomas} and the effects of confinement \cite{har2} etc., which can contribute effectively in obtaining the correct vacuum values of magnetic moments of octet baryons, and, for further analysis of magnetic moments in the presence of medium.

\end{document}